# Treatment of skin deseases in humans by a new method

Kozlov A.A.

*Laboratory of Developmental Biology, Tbilisi State University.*

*E-mail: alekskoz@rambler.ru*

Key words: skin covers, treatment of pathological processes, boric acid.

The original method of treatment of various pathological processes in human skin covers is described. The method has a brightly expressed differential action: destroying pathologic cells it does not render any influence on healthy cells. The examples of application of a method for treatment of some kinds of skin pathologies are given.

While pathological process develops in organism the immune system "turns on" to resist it. If the immune system is suppressed or the rate of pathology development is too high, the organism can not cope with it. It is possible to help the immune system by slowing down the rate of duplication of pathological cells.

In our former researches [1,2] it was shown, that the cell capable to proliferation, can divide only having received energy about 5 eV, transferred to it in one act. The basic role in this transfer is played with the photons of the appropriate power range formed in substance by a natural radiating background (nonequilibrium radiation). If into the water environment of cells the substances increasing probability of non-radiation dissipation of energy of excitation of molecules is inserted the number of the photons providing start of cell division will decrease therefore the number of divided cells for the given time interval will decrease also. Such substances, in particular, are: iodine, furaldehyde, alum, boric, salicylic and acetylsalicylic acids etc.

In that the given phenomenon takes place we were convinced of experiments with infusorians (*Colpoda sp.* and *Paramecia sp.*) [3]. Insertion of alum (**Al$_2$(SO$_4$)$_3$·K$_2$SO$_4$·24H$_2$O**) with concentration of $6 \times 10^{-7}$ M into a nutrient medium where the culture of infusorians developed, increased the average period of division of infusorians *Paramecia sp* for ~ 3.5 hours at the average period of division in the control of ~ 11 hours. Similar results were received with a boric acid (**$H_3BO_3$**).

After checking of this principle of cells division rate delay in laboratory conditions, the method was tested on treatment of some diseases of human skin covers, which are not carrying a system character.

As a substance, which is slowing down the rate of cell division, we used boric acid, which is also a good antiseptic. The pregnant solution of boric acid was applied to the best efficiency of influence. It was reached by imposing the wadded tampon, plentifully moistened by water, on which was rendered a gruel from boric acid on the struck area. From above the tampon was blocked by polythene. A bandage or court plaster fixed all this. After ending a procedure the rests of an acid were washed off by water. The number of seances of exposition and their duration can vary over a wide range. Usually two or three expositions of about 4-8 hours are enough.

The descriptions of 3 from more than 30 cases of practical use of this method on volunteers are given below. In total there were 8 different kinds of pathologies, the part of which was not diagnosed. The sex and age of the patients, description of disease, number of expositions and their duration are given. In the rest cases a complete recovery was also observed. No by-effects and relapses were detected.

1. M-55. Ring-shaped inflammation of a mole (5 mm in diameter) on forearm. 1 exposition, 4 hours. In one week the mole has disappeared. No traces were detected on its place.

2. M-45. A formation on a skin in area of a sural muscle. Round, light pink. Edges precise, equal. Does not act above skin. Border was felt to the touch. No pain, rarely itch. After washing - smallest peeling. The initial size was 1.5 - 2 cm in a diameter. In 1.5 months it has increased up to 4 - 5 cm. No fungus or other pathogenic microflora was revealed. Consultation of the dermatologists has not put an unequivocal diagnosis.

Under the initiative of the patient (a physician himself) the continuous exposition has been carried out within one week. Each day the tampon was moistened with water. After an exposition no trace of formation was revealed. No traces of influence on healthy skin.

3. M-60. Fungous defeat of fingers and toenails of both legs. Duration of disease - several years. Permanent itch between fingers. Medicamental treatment was carried out for a long time, periodically daily - to prevent itch. No positive result was obtained.

After the first exposition (8 hours) itch has stopped and had not been renewed since. For influence on toenail 2 additional expositions (after steaming out) for 8 hours were carried out. In 2 months a healthy toenail began to grow.

The preliminary studies showed also that cancer (early stages) can also be positively affected by this method.

The most important result of the described method of treatment (a priori, not expected), on our sight, is that fact, that the pregnant solution of boric acid, as it has appeared, destroying the pathologically changed cells, render absolutely no influence on healthy cells of skin. We are inclined to attribute this phenomenon to different permeability of cells membranes.

The recommendation: treatment of purulent inflammations previously wound to clear of from pus.